\documentclass[aps,pre,reprint,floatfix]{revtex4-2}

\usepackage[T1]{fontenc}
\usepackage[utf8]{inputenc}
\usepackage{lmodern}
\usepackage{microtype}

\usepackage[colorlinks=true,allcolors=blue]{hyperref}
\usepackage{xcolor}
\usepackage{ulem}

\usepackage{mathtools}
\usepackage{physics}
\usepackage{xspace}
\usepackage{amssymb}
\usepackage{calc}
\usepackage[linesnumbered,lined,boxed]{algorithm2e}

\newcommand*{\etal}{\textit{et al.}\xspace}

\newcommand*{\kT}{k_\text{B}T}
\newcommand*{\fa}{f_\text{a}}

\newcommand*{\Nf}{N_\text{fil}}

\newcommand*{\Rv}{R_\text{ves}}
\newcommand*{\Vv}{V_\text{ves}}
\newcommand*{\Vf}{V_\text{fil}}
\newcommand*{\bfil}{b_\text{fil}}
\newcommand*{\kFilBend}{\kappa_\text{fil}}

\newcommand*{\kVesBend}{\kappa_\text{ves}}

\newcommand*{\tR}{\tau_\text{rot}}
\newcommand*{\tC}{\tau_\text{coll}}

\newcommand*{\nCap}{n_\text{cap}}
\newcommand*{\Ushear}{U_\text{shear}}
\newcommand*{\Uint}{U_\text{int}}
\newcommand*{\Lrod}{L_\text{rod}}

\newcommand*{\lpDyn}{l_\text{p}^\text{COM}}

\newcommand*{\faSC}{f_\text{a}^\text{SC}}

\newcommand*{\FvK}{\text{FvK}}

\begin{document}

\title{Vesicle shape transformations driven by confined active filaments}

\author{Matthew S. E. Peterson}
\author{Aparna Baskaran}
\email{aparna@brandeis.edu}
\author{Michael F. Hagan}
\email{hagan@brandeis.edu}
\affiliation{
    Martin A. Fisher School of Physics,
    Brandeis University, Waltham, MA, 02453
}

\date{\today}

\begin{abstract}
In active matter systems, deformable boundaries provide a mechanism to organize internal active stresses and perform work on the external environment. To study a minimal model of such a system, we perform particle-based simulations of an elastic vesicle containing a collection of polar active filaments. The interplay between the active stress organization due to interparticle interactions and that due to the deformability of the confinement leads to a variety of filament spatiotemporal organizations that have not been observed in bulk systems or under rigid confinement, including highly-aligned rings and caps. In turn, these filament assemblies drive dramatic and tunable transformations of the vesicle shape and its dynamics. We present simple scaling models that reveal the mechanisms underlying these emergent behaviors and yield design principles for engineering active materials with targeted shape dynamics.
\end{abstract}

\maketitle


\section{Introduction}


Active matter encompasses systems whose microscopic constituents consume energy at the particle scale to produce forces and motion. Novel macroscale phenomena emerge in these systems when these forces collectively organize into mesoscale `active stresses'. Harnessing these active stresses to drive particular emergent behaviors could enable a new class of materials with life-like properties that would be impossible in traditional equilibrium materials. However, rationally designing active constituents to generate a desired emergent behavior requires identifying the principles that govern organization of mesoscopic active stresses. Similarly, many biological functions, such as cytoplasmic streaming, morphogenesis, and cell migration, are driven by active stresses that emerge from active components confined within a cell \cite{Sens2020, Ananthakrishnan2007,Leptin73, Miller2011, Polyakov2014,Besterman1983, Doherty2009, Maroudas-Sacks2020}. Understanding physical mechanisms that underlie these functions is a key goal of cellular biophysics.  

As a step toward these goals, this article describes a computational and theoretical study of a minimal model of active filaments enclosed within a deformable vesicle. We thereby identify a generic route to control self-organizing active stresses by enclosing active components with anisotropic shapes and/or internal degrees of freedom within deformable confining boundaries. Our results demonstrate that coupling between boundary deformations and the assembly of internal active components leads to a positive feedback capable of driving diverse nonlinear transformations of morphology and dynamics. 

The field of active matter has identified two key mechanisms that control self-organization of active stresses: (1) anisotropic interactions between active components that realign forces, and (2) confining boundaries. For example, interactions between self-propelled particles that drive interparticle alignment result in bands or flocks~\cite{Schaller2010, Bricard2013}, changing the length and stiffness of active polymers leads to dramatic reorganization of active stresses~\cite{Senoussi2019, Strbing2020}, and confining active particles leads to system-spanning effects such as spontaneous flow~\cite{Edwards2009, Woodhouse2012, Wioland2013, Ravnik2013, Lushi2014, Wioland2016, Wu2017, Varghese2020}.  Furthermore, deformable confining boundaries enable non-equilibrium boundary fluctuations~\cite{Li2019a, Vutukuri2020, Keber2014, Takatori2020, Hughes2020, Miller2018}, including elongated tendrils and bolas~\cite{Vutukuri2020}. The latter results suggest that flexibility is a key characteristic of a confining boundary, as it allows shape transformations, sensing and response to environmental cues, and performing work on the surroundings. Achieving such capabilities is critical to leverage minimal bio-derived experimental systems (e.g.~\cite{Elbaum1996, Fygenson1997, Keber2014}) to engineer controlled shape transformations. However, little is known about the behaviors that may arise when these two active stress organization modes are combined. In particular, most existing theoretical and computational studies have focused on rigid boundaries~\cite{Norton2018, Henkes2018, Shankar2017, Alaimo2017}, isotropic active particles~\cite{Paoluzzi2016, Li2019a, Vutukuri2020, Takatori2020, Wang2019, Sknepnek2015, Bruss2017}, or have been in 2D~\cite{Abaurrea-Velasco2019, Quillen2020, Vliegenthart2020}.

In this work, we use Langevin dynamics simulations of polar self-propelled semiflexible filaments confined within 3D flexible vesicles to study the combination of active interparticle alignment interactions and 3D deformable confinement. Our simulations show that interplay between these two organization modes leads to a positive feedback, in which active forces drive boundary deformation while passive stresses from the boundary guide and reinforce self-organization of the internal active stresses. This leads to a rich variety of steady-state behaviors that have not been observed in bulk systems or under rigid confinement, including highly-aligned rings, and caps that have tunable self-limited sizes, number, and symmetry. Each filament organization drives a characteristic large-scale vesicle shape transformation that can be selected by varying parameters such as filament length, density, and flexibility. Asymmetric states lead to net vesicle motion, consistent with the recent experiments which find that enclosing self-propelled particles, such as bacteria, in droplets can lead to collective motility~\cite{Rajabi2020}. We present simple scaling analyses that reveal how the feedback between vesicle geometry and filament organization drives and stabilizes these emergent behaviors. The applicability of these scaling arguments suggests that these behaviors arise generically due to feedback between vesicle elasticity and active filament organization, independent of the specific model.

Understanding the fundamental mechanisms that govern this coupling between self-organization of active stress and deformations of a flexible boundary will establish design principles for soft robotics, artificial cells, or other advanced materials that mimic the capabilities of living organisms. From a biological perspective, our minimal model is not intended to directly describe the cytoskeleton---in this model, active filament propulsion enters in the overdamped limit without long-ranged hydrodynamic coupling. However, the generic feedback between vesicle elasticity and active filament organization identified in our simulations and scaling arguments may elucidate how mesoscopic active stresses in the cellular cytoskeleton drive large scale cellular shape transformations that underlie essential biological functions such as motility~\cite{Sens2020, Ananthakrishnan2007}, division~\cite{Leptin73, Miller2011, Polyakov2014}, and endo-/exocytosis~\cite{Besterman1983, Doherty2009}. 


\section{Methods}

We simulate a system of $\Nf$ active filaments confined within an elastic vesicle, which has radius $\Rv$ in its undeformed state. We represent active filaments using the model in Joshi, \etal~\cite{Joshi2019}---modified so that the active forcing is polar rather than nematic---in which each filament is a nearly-inextensible, semiflexible chain of $M$ beads of diameter $\sigma$~\cite{Winkler2020}. Bonded beads interact through an expanded FENE potential~\cite{Kremer1990}, while non-bonded beads interact through a purely repulsive expanded Weeks-Chandler-Andersen (eWCA) potential~\cite{Weeks1971} with strength $\epsilon$. The equilibrium bond length is set to $\bfil=\sigma/2$ to minimize surface roughness between interacting filaments, thereby preventing filaments from interlocking at high density~\cite{Elgeti2013, Isele-Holder2015, Isele-Holder2016, Duman2018,Chelakkot2020}. The filaments are made semiflexible with bending rigidity $\kFilBend$ through a harmonic angle potential applied to each set of three consecutive beads along the chain. Since our model is not intended to describe any specific biofilament system, we incorporate activity in a minimal manner---a polar active force  of magnitude $\fa$ acts on each bead, in a direction tangent to the filament and toward the filament head. The filament volume fraction in the undeformed state is given by $\phi=\Nf \Vf / \Vv$, where $\Vf = \pi \sigma^3 / 6 + (M - 1) \pi b \sigma^2 / 4$ is the approximate volume of a single filament---accounting for the overlap of bonded monomers---and $\Vv = 4 \pi \Rv^3 / 3$ is the nominal volume of the vesicle. Since the mesh topology is conserved in our simulations, we model an elastic vesicle. In subsequent work we plan to consider the effects of fluidizing the vesicle and imposing area or volume constraints. 

We simulate the coupled Langevin equations for the filament and vesicle bead dynamics using LAMMPS~\cite{Plimpton1995}, modified to include the active force. We neglect long-ranged hydrodynamic interactions for this system of high filament density; we will investigate their effect in a future study. We have set units such that the mass of all beads is $m = 1$, and energies, lengths, and time are respectively in units of $\kT$, $\sigma$, and $\tau = \sqrt{m \sigma^2 / \epsilon}$. The friction constant is set to $\gamma = 1/\tau$. For additional model details, see the Supplemental Materials~\cite{SIref}.


\section{Results and discussion}

\begin{figure}
    \centering
    \includegraphics{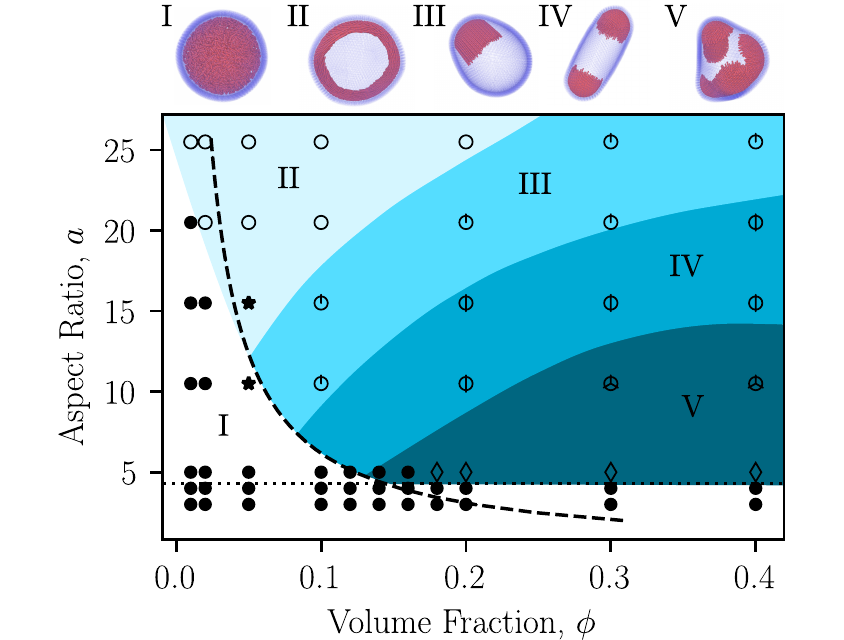}
    \caption{Snapshots illustrating steady-state configurations of the vesicle and enclosed active filaments as a function of filament aspect ratio $a$ and initial volume fraction $\phi$. See SI Movie 1 for animations of the corresponding simulations~\cite{SIref}. The marked regions of parameter space indicate the typical vesicle conformation: \textit{(I)} spherical, \textit{(II)} oblate, \textit{(III)} polar-prolate, \textit{(IV)} apolar-prolate, and \textit{(V)} polyhedral. The symbols associate the conformation with the internal filament organization: homogeneous throughout the bulk or on the surface, with no vesicle deformation ($\bullet$); transient clusters and/or bands, with oblate vesicle shapes ($\star$); stable polar rings ($\circ$); stable caps ($\circ$, with a number of intersecting lines equal to the median number of caps); and dynamic caps ($\Diamond$). The dashed line shows the transition to aligned states predicted from the competition between the characteristic collision and reorientation timescales ($\phi=(\pi/4)^2/a$) described in the text, and the horizontal dotted line indicates the approximate threshold aspect ratio for the filaments to be in the strong confinement limit. Other parameters are filament bending modulus $\kFilBend=10^4$ and vesicle radius $\Rv=25$.}
    \label{fig:density-length-states}
\end{figure}

\begin{figure}
    \centering
    \includegraphics{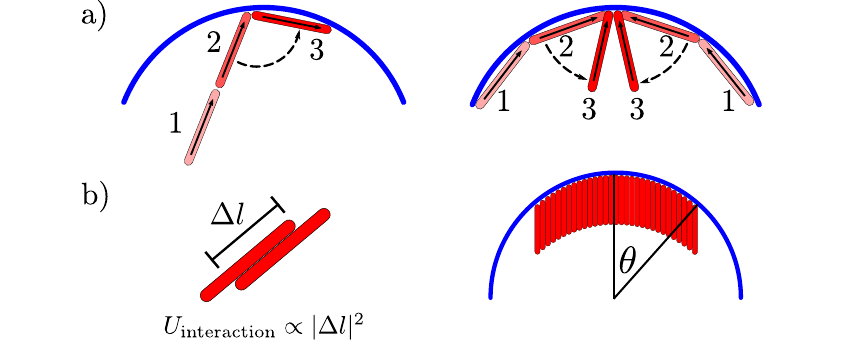}
    \caption{\textbf{a)} The onset of ring and cap formation is determined by a competition of timescales: the timescale associated with rotations parallel to the vesicle (left) and the timescale associated with collisions that tend to orient the filaments perpendicular to the vesicle (right). \textbf{b)} Schematic of the theory for the number of caps (Eqs.~\eqref{eq:fShear} and \eqref{eq:nCaps}). We assume an activity-induced effective attractive interaction that is quadratic in the rod-rod contact length $\Delta l$ (left). The cap is assumed circular, with size parameterized by the angle $\theta$ between the cap center and edge. Vesicle curvature leads to a shearing of rods within the cap (right).}
    \label{fig:timescale-competition}
\end{figure}

\subsection{Simulation results}
To discover the steady-state conformations that arise due to coupling between active propulsion and elasticity, we have performed simulations over a wide range of control parameters (Fig.~\ref{fig:density-length-states}, Fig.~\ref{fig:activity-density-states}, Fig.~\ref{fig:stiffness-activity-states}, and Fig.~\ref{fig:dihedral-density-states}): the volume fraction of filaments in the vesicle $\phi \in [0.01, 0.4]$, filament aspect ratio $a = 1 + (M - 1)(\bfil / \sigma) \in [3, 25.5]$, active propulsion strength $\fa \in [0, 10]$, filament stiffness $\kFilBend \in [10^2, 10^4]$, and vesicle rigidity $\kVesBend \in [10^2, 10^4]$.

Fig.~\ref{fig:density-length-states} shows the steady-states as a function of filament volume fraction and aspect ratio for moderate activity $\fa=8$. At this activity and vesicle size, for aspect ratios $a \gtrsim (8\pi\Rv/\fa)^{1/3}\approx 4.3$ the system is in the \textit{strong confinement limit}: because the persistence length $\lpDyn \propto \fa a^3$ of the filament center-of-mass motion is larger than the vesicle size $\lpDyn > 2\Rv$, most filaments are found on the vesicle surface at all times~\cite{Fily2014} (see SI Sec.~A~\cite{SIref}).

Under these conditions we can classify the steady-state vesicle conformations into several categories: \textit{(I)} spherical,  \textit{(II)} oblate, \textit{(III)} polar-prolate, \textit{(IV)} apolar-prolate, and \textit{(V)} polyhedral. These vesicle configurations are tightly coupled to the spatiotemporal organization of the filaments within, as follows.

\textit{(I)}: Spherical vesicle shapes arise at low filament volume fractions and aspect ratios. Under these conditions, filament-filament collisions are rare and inter-filament aligning forces are weak~\cite{Peruani2006, Peruani2012, Baskaran2008, Baskaran2008b, McCandlish2012, Ginelli2010, Baer2020}. Thus, filament positions and orientations are homogeneous (throughout the vesicle interior below strong confinement, or on the vesicle surface above strong confinement), leading to little deformation of the vesicle. 

\textit{(II)}: For low volume fraction but high aspect ratios, such that the filament length $L = a\sigma$ is comparable to the unperturbed vesicle radius, $L \sim \Rv$, the vesicle deforms into oblate spheroid conformations. This transition is driven by the filaments organizing into a stable polar band, which deforms the vesicle along a geodesic. This filament arrangement closely resembles the polar bands observed on the surface of rigid spheres for active particles with polar propulsion and polar interparticle alignment interactions~\cite{Sknepnek2015}, which arise due to topological requirements for a surface-constrained polarization field~\cite{Shankar2017}. However, note that such polar bands would be unstable in our system if the confining geometry was a rigid sphere because the filament-filament interactions in our system are nematic (head-tail symmetric)~\cite{Bertin2013,Ngo2014}. The finite deformability of the vesicle is essential to stabilize this configuration---active forces due to the polar band force the vesicle into an oblate shape, which in turn provides a restoring force to stabilize filament alignment within the band. In support of this conclusion, simulations on infinitely rigid vesicles did not exhibit stable polar bands (see Fig.~\ref{fig:dihedral-density-states}b and SI movie 6). Thus, this configuration provides a concrete example of how feedback between passive stresses and self-organization of active stresses can generate steady states that would be otherwise disallowed by symmetry.

\textit{(III-IV)}: For intermediate volume fractions and aspect ratios, the vesicle deforms into a prolate spheroid. These prolate vesicle conformations can be further classified by their motion, either polar \textit{(III)} or apolar \textit{(IV)}. Further increasing the volume fraction or decreasing the aspect ratio leads to polyhedral conformations, \textit{(V)}. States \textit{(III-V)} all result from filaments assembling into crystalline caps in which the rods are highly aligned and perpendicular to the vesicle surface. Interestingly, the caps are `self-limited' in that their typical size decreases with decreasing aspect ratio, but is roughly independent of the total number of filaments $\Nf$ in the vesicle. Increasing $\Nf$ at fixed aspect ratio increases the number of caps; we observe up to 12 caps for the finite vesicle size that we consider (Fig.~\ref{fig:cap-count-estimate}). Further, caps drive local curvature of the vesicle, leading to elasticity-mediated cap-cap repulsions which favor symmetric arrangements of caps. Thus, the vesicle morphology can be sensitively tuned by controlling filament aspect ratio and density to achieve a specific number of caps. The polar-prolate \textit{(III)}, apolar-prolate \textit{(IV)}, and polyhedral states \textit{(V)} respectively have 1, 2, and $\ge 3$ caps. Generally, states with two or more caps do not exhibit directed motion. However, for enough caps in the vesicle (typically more than 3), the caps themselves can become motile, and collide with, merge with, and split from other caps (see below).

\subsection{Mechanisms underlying stress organization and deformation}
To understand how these conformations are governed by the interplay between propulsion-induced aligning forces, vesicle deformability, and vesicle curvature, we develop simple scaling estimates for the timescales and forces that govern filament alignment and interactions with the vesicle. First, we consider the transition between undeformed spherical vesicle states characterized by unaligned or weakly aligned filaments as in state \textit{(I)}, to the highly deformed oblate, prolate, and polyhedral vesicle shapes of states \textit{(II-V)}. Our simulations demonstrate that such significant vesicle shape deformations occur when filament-filament interactions mediate the organization of ordered structures either in the plane of the vesicle or orthogonal to it.

\paragraph{The onset of filament assembly:}
The onset of this transition can be understood by considering a competition between two characteristic timescales that respectively govern collision-induced filament-vesicle alignment and filament-filament alignment (see Fig.~\ref{fig:timescale-competition}). Filament-vesicle collisions, which tend to reorient filaments parallel to the surface~\cite{Wensink2008, Bechinger2016}, have a characteristic timescale $\tR \sim L / v_0$
\footnote{The ability of collisions of self-propelled particles on a surface to drive formation of smectic layers is supported by a recent observation in bacterial colonies growing on flat surfaces, in which bacteria form `rosettes' with the rod-like bacteria oriented perpendicular to the surface~\cite{Meacock2020}. However, in contrast to the self-limited caps in our system, the bacterial rosettes do not exhibit a preferred size because they are on a flat boundary.},
with $v_0 = \fa / \gamma$ the filament self-propulsion velocity. We can estimate the timescale for filament interactions by considering filament-filament pairwise collisions whose timescale is given by $\tC \sim \sigma / v_0 \phi$ (see SI~\cite{SIref}). Thus, deformed vesicle states will arise when $\tC < \tR$ or equivalently $a \phi > c$, where $c\cong (\pi / 4)^2$ is independent of activity and filament length (see~\cite{SIref}). This defines a boundary separating highly deformed states of the vesicle from the undeformed spherical states (the dashed line in Fig.~\ref{fig:density-length-states}).

\begin{figure}
    \centering
    \includegraphics{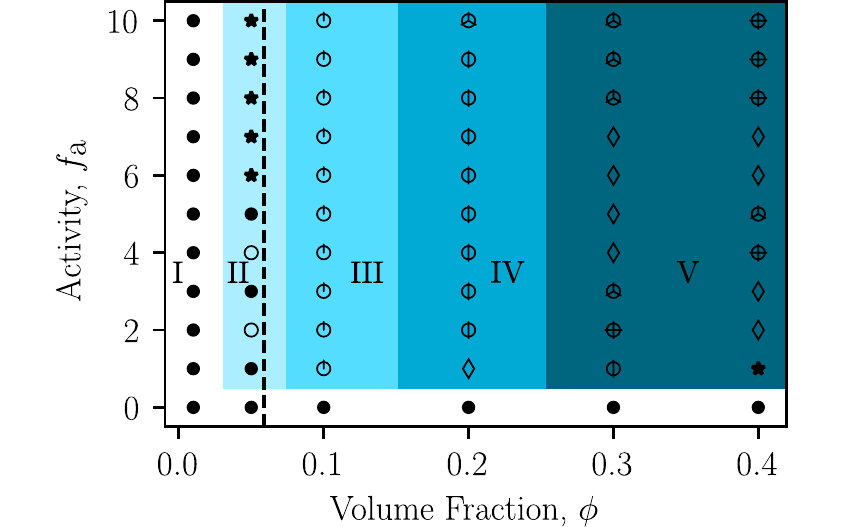}
    \caption{Steady-state configurations as a function of $\phi$ and active force $\fa$. The marked regions are defined as in Fig.~\ref{fig:density-length-states}. The dashed line shows the transition to aligned states predicted by the timescale competition, which is independent of $\fa$. Other parameters are $a=10.5$ and $\kFilBend=10^4$. See SI Movie 2 for corresponding animations.}
    \label{fig:activity-density-states}
\end{figure}

Notably, the active force drops out of this argument because both collision and reorientation times are $\propto \fa$. Thus, the theory predicts that the emergence of deformed vesicle states is independent of activity of the enclosed filaments (above a threshold activity). As a test of this prediction, Fig.~\ref{fig:activity-density-states} shows the steady-states as a function of $\phi$ and $\fa$ for fixed aspect ratio $a=10.5$. Indeed, formation of large deformations does not depend on activity, with non-spherical shapes forming for $\phi \geq c/a \approx 0.06$ (as predicted by the above timescale argument) for all $\fa > 0$ that we considered.

This simple theoretical picture gives a predictive principle, in terms of properties of the active filaments, for when vesicle shape transformations occur. However, the theory assumes the strong activity, long filament limit and thus neglects thermal noise. Below a threshold activity ($\fa\lesssim 1$ in our units) the vesicle will not deform because filament organization is destroyed by thermal fluctuations. Also, cap formation (and thus vesicle shape transitions) do not occur when the filaments are below the strong confinement limit discussed above ($a \lesssim 4.3$ for the parameters of Fig.~\ref{fig:density-length-states}, shown as a dotted line).

\begin{figure}
    \centering
    \includegraphics{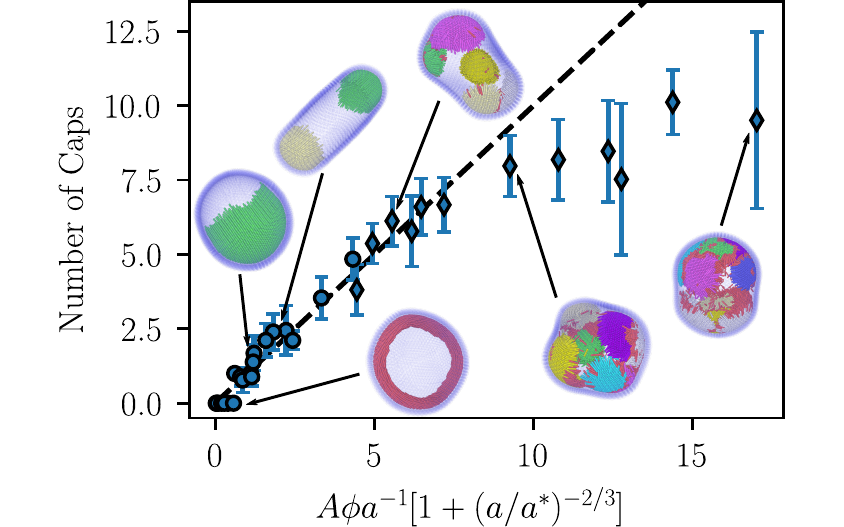}
    \caption{The number of caps measured in simulations (symbols) compared to the theory (Eqs.~\eqref{eq:fShear} and \eqref{eq:nCaps}, dashed line), with $A=4 \Rv / \sqrt{3} \pi \sigma$ and $a^* \approx 130$ (chosen by eye). Diamonds indicate dynamic cap states. Note that the number of caps in the simulation results is likely under counted for the dynamic states due to the caps' motility. The simulation data is the same as in Fig.~\ref{fig:density-length-states}. Active filaments are colored by which cap they belong to for visual clarity.}
    \label{fig:cap-count-estimate}
\end{figure}

\paragraph{Cap morphologies:}
We can derive further insight into shape transformations by considering the system in the strongly deformed regime with polyhedral shapes. The defining characteristic underlying these states is filament assembly into well-ordered caps.  Most cap states are relatively static, with occasional association/dissociation of individual rods (See SI Movie 4), except for the parameters that lead to the highly dynamic, reconfiguring caps discussed below. In a static steady state, the active and elastic forces must balance. In particular, the dense crystalline nature of caps arises because the active force and the presence of the vesicle surface leads to an effective attractive interaction between nearby filaments. This attraction drives radial growth of a cap, since filaments on the cap periphery have fewer neighbors, leading to an effective interfacial tension. This effect is both reinforced by and competes with vesicle elasticity. The active force of small caps drives vesicle deformations whose local curvature enhances effective filament-filament attractions. However, as the cap grows in radius, vesicle curvature drives an effective shear of filaments (see Fig.~\ref{fig:timescale-competition}b) that reduces rod-rod overlaps and thus opposes the active force.

We describe this competition by constructing an effective `free energy' whose gradients correspond to the active and passive forces (Fig.~\ref{fig:timescale-competition}b). Since the active force favors rods to align in a smectic layer, the shear due to vesicle curvature imposes an `energy' cost of $\Ushear(\theta) =  \nCap 2 \pi \Rv^2 \frac{G}{2} \left[\cos\theta + \sec \theta -2 \right]$, with $\theta$ the angle subtended by the cap on the vesicle surface, $\nCap$ the number of caps, and a `shear modulus' $G \thicksim \fa$ (but independent of $\Lrod$)~\cite{SIref}. In the strongly deformed region the caps are roughly circular, so the interfacial energy is given by $\Uint(\theta) = \nCap 2 \pi \Rv \gamma \sin \theta$, with the `interfacial tension' $\lambda \thicksim \Lrod \fa$ accounting for the diminished interactions at the cap boundary. This results in a free energy as a function of cap size~\cite{SIref}:
\begin{equation}
    f(\theta) = \frac{1}{1 - \cos\theta} \left[
        \frac{1}{2}(\cos \theta + \sec \theta - 2)
        + \zeta \sin \theta
    \right]
\label{eq:fShear}
\end{equation}
where $\zeta = G/\gamma \Rv \sim L / \Rv$ is given by the balance between the effective interfacial tension and shear modulus, and should be linear in filament length but roughly independent of $\fa$ since both of these effects are driven by activity.

Minimizing this per-filament free energy yields an optimal $\theta$~\cite{Hagan2021, Yu2016} corresponding to the self-limited cap size. Assuming that we are well above the onset of cap formation so that essentially all filaments are in caps,
\begin{equation}
    \nCap \propto \phi a^{-1} [1 + (a / a^*)^{-2/3}]
\label{eq:nCaps}
\end{equation}
where $a^* \propto \Rv / \sigma$ is an adjustable parameter that may depend on activity. This expression holds provided $a \ll a^*$. For the data in Fig.~\ref{fig:density-length-states}, we obtain $a^* \approx 130$, leading to the dashed line shown in Fig.~\ref{fig:cap-count-estimate}.

Except for states with many ($\nCap \gtrsim 7)$ motile caps, there is close agreement between the observed and predicted $\nCap$. Above this threshold our cap-counting algorithm likely under counts $\nCap$, since different caps are often adjacent and interacting. Further, the prediction of Eqs.~\eqref{eq:fShear} and \eqref{eq:nCaps} that the self-limited cap size is independent of activity is consistent with observations at different $\fa$ (see Fig.~\ref{fig:activity-density-states}). The motile cap states appear to arise when the curved vesicle geometry forces interactions between the inward-facing ends of adjacent caps. Such interactions occur above a threshold number and aspect ratio of filaments, given by $\Nf \gtrsim C (1 - a \sigma / \Rv)^2$, where $C$ is a constant (see SI Sec.~D~\cite{SIref}).

We note that the geometric factors governing the self-limited cap size parallel those in a recently studied \textit{equilibrium} system of rigid filaments end-adsorbed onto a rigid spherical nanoparticle, which self-assemble due to direct pairwise inter-filament attractions \cite{Yu2016}.  However, in the present system, the effective filament-filament interactions and vesicle geometry are many-body and emergent in that they arise due to feedback between non-equilibrium active forces and vesicle deformations.


\subsection{Effect of filament and vesicle rigidity}

\begin{figure}
    \centering
    \includegraphics{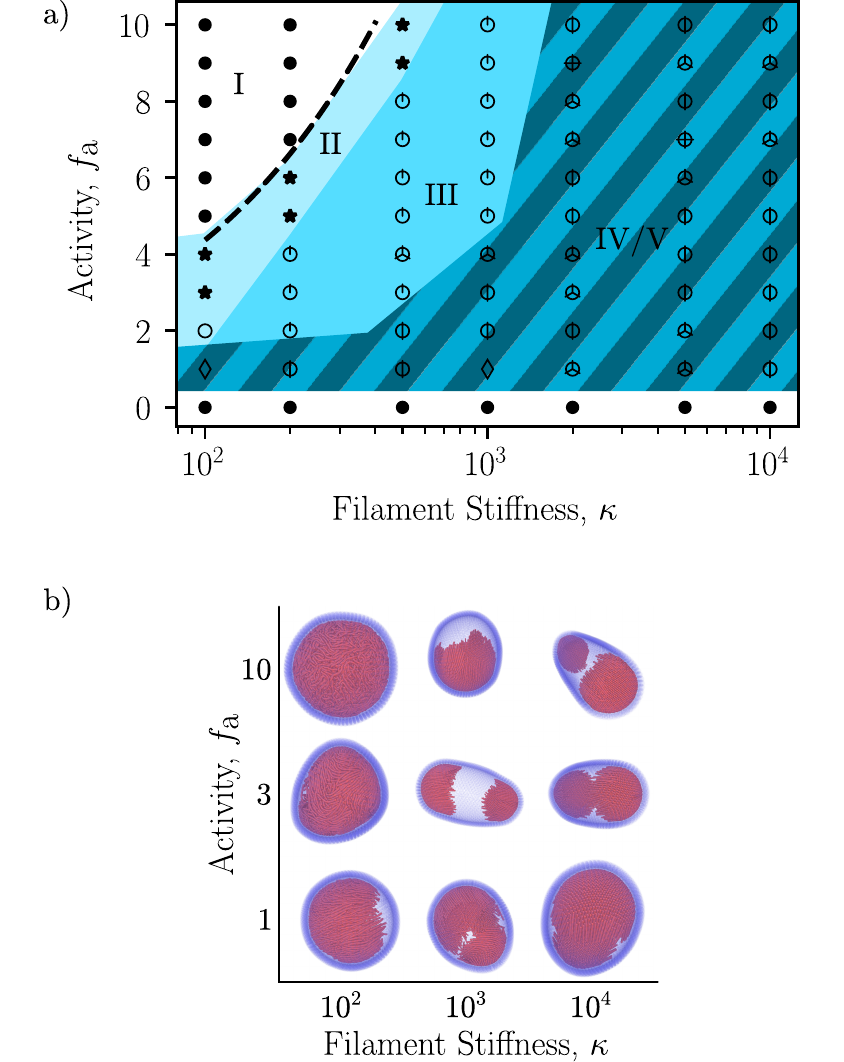}
    \caption{\textbf{a)} Vesicle conformations and filament organizations as a function of filament rigidity $\kFilBend$ and active force strength $\fa$, for volume fraction $\phi = 0.2$ and filament aspect ratio $a = 10.5$. For a given filament stiffness, increasing activity reduces the number of caps until an upper-threshold activity value $\faSC$, beyond which the system transitions into an undeformed state. As described in SI section E, this transition occurs because activity renormalizes the filament bending modulus to smaller values\cite{Joshi2019}, thus reducing filament alignment interactions and causing the system to leave the strong confinement limit. The dashed line shows the prediction for $\faSC$ given by Eq.~S35. Note that there is no adjustable parameter. In the rigid rod limit ($\kFilBend > 10^3$) all non-zero active force values that we simulated led to cap formation. \textbf{b)} Selected snapshots of states shown in (a). Animations of these states can be found in SI Movie 3.}
    \label{fig:stiffness-activity-states}
\end{figure}

Thus far, we have focused on the interplay between activity and vesicle deformability by performing simulations in the limit of rigid rods, $\kFilBend=10^4$, and high (but finite) vesicle rigidity $\kVesBend=5 \times 10^3$. We now briefly discuss the effect of allowing for finite filament and vesicle flexibility.

Fig.~\ref{fig:stiffness-activity-states} shows the vesicle conformation and filament organization states as a function of filament bending modulus and activity, for fixed filament volume fraction $\phi=0.2$. We see that for finite filament flexibility, the transition to aligned ring and cap states is suppressed \textit{above} a threshold activity, which decreases with decreasing $\kFilBend$.

This result can be understood as follows. On generic grounds, decreasing the filament rigidity will reduce the tendency for filaments to align and thus impede the formation of aligned rings and caps. For filament stiffness values well below the rigid rod limit, the process by which caps and rings form is more complicated than considered previously. The upper-threshold activity for filament organization can be, at least in part, explained by the observation that activity renormalizes filament rigidity to smaller values according to $\kFilBend^\text{eff} \cong \kFilBend/\left(1+\fa^2\right)$~\cite{Joshi2019}. Interactions between flexible active agents is such that the active energy preferentially dissipates into bend modes, effectively increasing filament flexibility and therefore suppressing filament alignment. In particular, the upper-threshold activity corresponds to the point when the activity-renormalized flexibility of filaments causes the system to leave the strong confinement limit. This occurs for $\fa \gtrsim C \kFilBend^{3/5}$, where $C = (8 \pi \Rv)^{-1/5}$ (see SI~Sec.~D for details), which is shown as the dashed line in Fig.~\ref{fig:stiffness-activity-states}.


\begin{figure}
    \centering
    \includegraphics{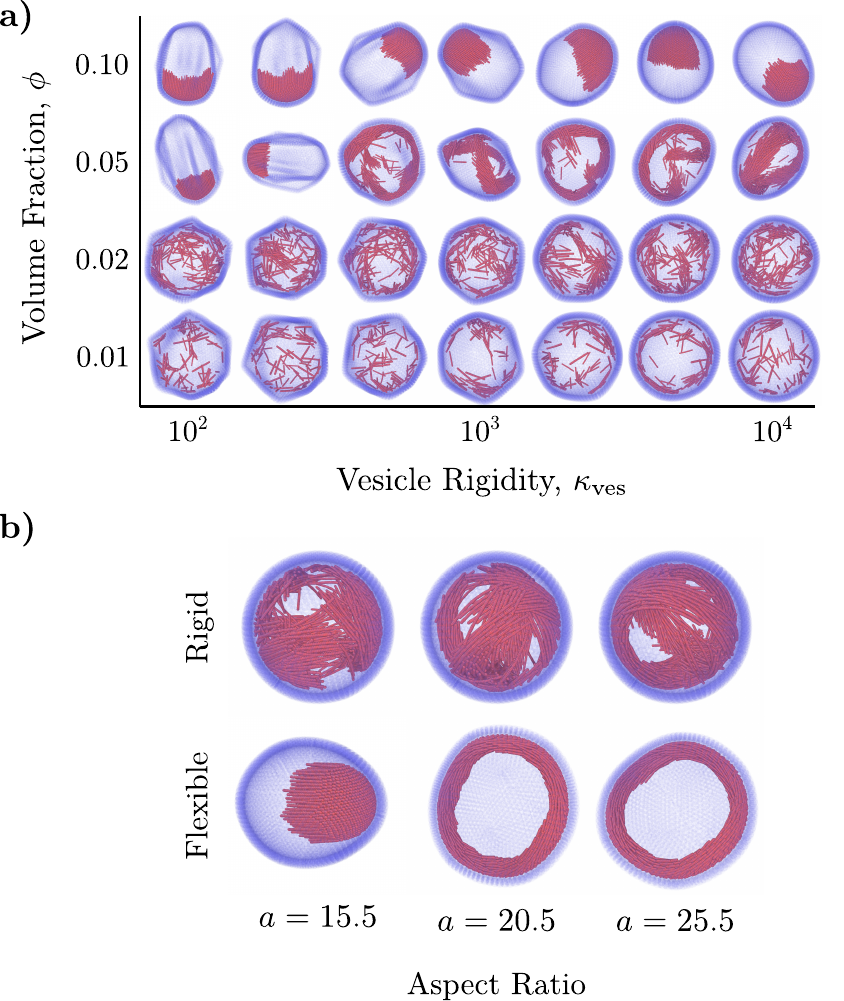}
    \caption{\textbf{a)} Simulation snapshots illustrating vesicle conformations and filament organizations as a function of filament volume fraction and vesicle rigidity. As the vesicle rigidity is reduced below a critical value (corresponding to the critical Föppl–von Kármán number $\FvK\approx154$ \cite{Lidmar2003}), the vesicle undergoes a buckling transition leading to the formation of facets. While we observe most of the same classes of filament self-organization in faceted and round vesicles, polar bands trace a dynamic path between vertices in faceted vesicles, while they trace a stable geodesic in round vesicles. Animations of these states can be found in SI movie 5. \textbf{b)} Comparisons between flexible and rigid vesicles as a function of filament aspect ratio, with other parameters set to $\fa = 8$, $\phi = 0.1$, and $\kFilBend = 10^4$. In contrast to flexible vesicles, rigid vesicles do not allow for the formation of stable caps or rings. When polar rings do form in rigid vesicles, they are transient---continuously breaking and reforming over the course of the trajectory. Animations of this comparison can be found in SI movie 6.}
    \label{fig:dihedral-density-states}
\end{figure}

Fig.~\ref{fig:dihedral-density-states}a shows the conformations obtained by varying the vesicle rigidity $\kVesBend$ and filament volume fraction $\phi$, while fixing the filament rigidity $\kFilBend = 10^4$ and active force $\fa = 8$. The most striking effect of reducing the vesicle rigidity is that it drives a faceting transition when the the Föppl–von Kármán number, $\FvK = Y \Rv^2 / \kappa$ where $Y$ is the Young's modulus of the vesicle and $\kappa=\sqrt{3}\kVesBend$ is the bending modulus~\cite{Gompper1996}, is increased above a critical value, $\FvK \gtrsim 154$. This is an equilibrium property of an elastic vesicle, independent of the active filaments~\cite{Lidmar2003}. Our results indicate that faceting does not qualitatively change the formation of caps, but that caps form at slightly lower filament volume fraction for reduced vesicle bending modulus. This could be anticipated from the theoretical arguments described above, since reducing the bending modulus allows filaments’ active forces to further deform the vesicle, leading to a smaller local radius of curvature in the vicinity of a cap.  More interestingly, the facets appear to destabilize the polar bands and rings. For round vesicles (with bending modulus such that $\FvK<154$), a stable ring forms along a geodesic. In contrast, in faceted vesicles at the same activity and filament volume fraction, rings or bands tend to form paths that connect facet vertices. The bending of the ring path imposed by the facet connectivity destabilizes the ring, causing it to transiently break and reform (similar to the transient band state described above). This behavior suggests that it will be interesting to explore the possibility of coupling between vesicle faceting and filament organization in a future work.

Fig.~\ref{fig:dihedral-density-states}b compares configurations observed with a flexible vesicle ($\kVesBend = 5 \times 10^3$) and a rigid vesicle ($\kVesBend \to \infty$) for $a \in [15.5, 25.5]$, $\phi = 0.10$, $\fa = 8$, and $\kFilBend = 10^4$. While the flexible vesicle exhibits stable polar rings and single caps at these parameters (Fig.~\ref{fig:density-length-states}), the rigid vesicle system is unable to form the single-cap state, and only exhibits transient polar rings, which continuously break apart and reform as the simulation progresses. These results emphasize the importance of the feedback between active stress organization and vesicle deformation, which allows for stable states that are otherwise inaccessible under rigid confinement.


\section{Conclusions}

This work demonstrates that confining active filaments within a deformable vesicle leads to multiple transformations of the vesicle shape and motility, which can be precisely tuned by control parameters. The feedback enabled by coupling deformable boundaries with anisotropic particles significantly enriches the available modes of self-organization. While the self-limited caps are the most striking class of such behaviors, the stable polar bands for particles with nematic interactions provides a clear example of how boundary deformations can stabilize novel states. Notably, both of these classes of behaviors arise due to a spontaneous symmetry breaking of the initially spherical boundary.

These results have implications for future experiments on active materials constructed from anisotropic particles confined within deformable boundaries. In particular, the transitions can be controlled by tuning parameters that are readily accessible in experiments---filament length, flexibility, and volume fraction. In contrast, activity is a complicated function of motor properties and ATP in bio-derived systems~\cite{Lemma2020, Chandrakar2018}. Thus, our computational results suggest strategies to engineer active vesicles with designable shapes and dynamics, and other capabilities resembling those of living cells. Furthermore, our theoretical analysis identifies the mechanisms that underlie these emergent morphologies by revealing how filament-filament interactions and vesicle deformations couple to spatiotemporally organize stress. This provides a model-independent roadmap for exploring additional classes of emergent functionalities in parameter regimes beyond the scope of the present work, including highly deformable fluidized vesicles and other symmetries of activity.

\begin{acknowledgments}
We acknowledge support from NSF DMR-1855914 and the Brandeis Center for Bioinspired Soft Materials, an NSF MRSEC (DMR-2011846). We also acknowledge computational support from NSF XSEDE computing resources allocation TG-MCB090163 (Stampede and Comet) and the Brandeis HPCC which is partially supported by DMR-MRSEC 2011486. We also acknowledge the KITP Active20 program, during which some of these ideas were developed, which is supported in part by the National Science Foundation under Grant No. NSF PHY-1748958.
\end{acknowledgments}

\textbf{Author contributions:} MSEP, AB, and MFH designed the research; MSEP performed the computational modeling; MSEP, AB, and MFH performed the theoretical modeling; MSEP analyzed the data; and MSEP, AB, and MFH wrote the paper.

\textbf{Competing interests:} The authors declare that they have no competing interests.

\bibliographystyle{apsrev4-1}
\bibliography{references}







\end{document}


\onecolumngrid

\title{Supplementary information for Vesicle shape transformations driven by confined active filaments}

\author{Matthew S. E. Peterson}
\author{Aparna Baskaran}
\email{aparna@brandeis.edu}
\author{Michael F. Hagan}
\email{hagan@brandeis.edu}
\affiliation{
    Martin A. Fisher School of Physics,
    Brandeis University, Waltham, MA, 02453
}

\date{\today}

\maketitle

\onecolumngrid

\tableofcontents

\newpage

\section*{Supplementary Figures}

\begin{figure}[h]
    \centering
    \includegraphics{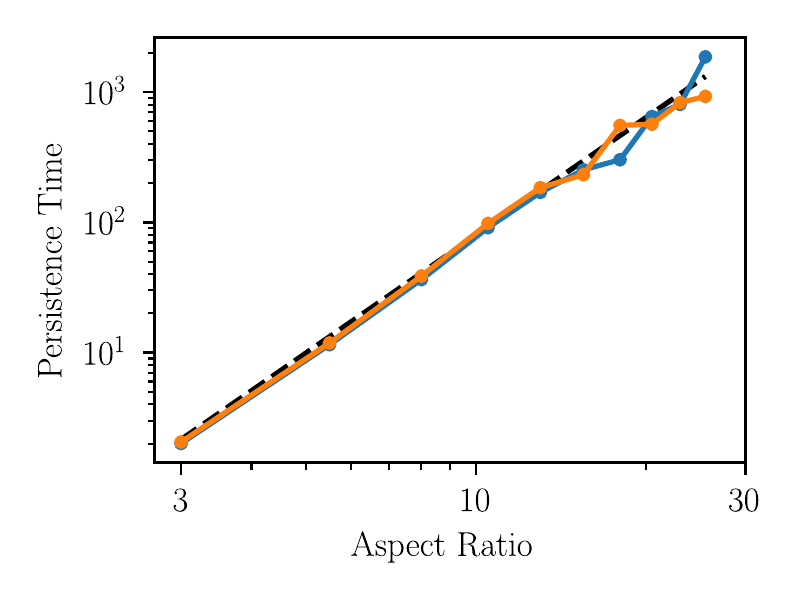}
    \caption{The filament orientation persistence time scales with aspect ratio as $\tCorr \propto a^3$ in the rigid rod limit ($\kFilBend = 10^4$). The blue dots and orange dots are data from simulations with no interfilament interactions with the active force turned off and on, respectively, showing that activity does not affect the persistence time. The dashed line follows $\tCorr = a^3 / 4\pi\gamma$.}
    \label{fig:persistence-time}
\end{figure}


\begin{figure}[h]
    \centering
    \includegraphics{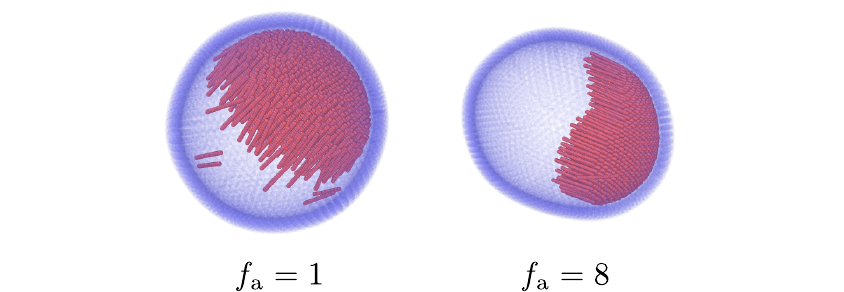}
    \caption{Caps that form at lower active force strengths (left) tend to be more ragged and drive less vesicle deformation than those that form at larger active force strengths (right). For both snapshots, $\phi = 0.2$, $a = 10.5$, and $\kFilBend = 10^4$.}
    \label{fig:ragged_caps}
\end{figure}

\begin{figure}[h]
    \centering
    \includegraphics{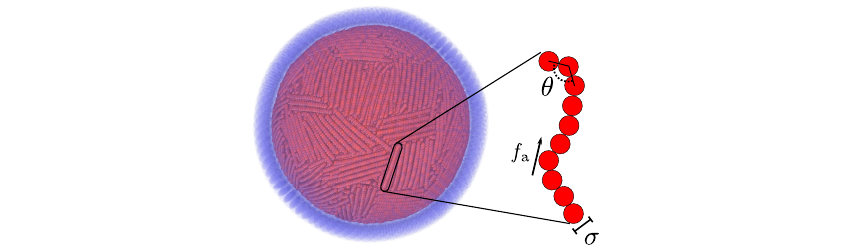}
    \caption{The vesicle is constructed from a triangulated mesh of $\Nv$ beads, and contains $\Nf$ polar active filaments. Each filament consists of $M$ bonded monomers of diameter $\sigma$. A propulsion force with magnitude $\fa$ is applied to each monomer in the direction of the local filament tangent, and the filament bending modulus is $\kFilBend$.}
    \label{fig:model}
\end{figure}

\clearpage

\section*{Supplementary Text}

\subsection{Filament persistence time}

In this section we compute the persistence length of the center of mass motion of the active filaments, $\lpDyn$, to determine the regions in parameter space that correspond to the strong confinement limit. Under strong confinement, almost all rods are on the vesicle surface at all times, while below this limit rods are found throughout the vesicle interior. Strong confinement occurs when the filament motion persistence length is larger than the vesicle size, i.e., $\lpDyn \gtrsim 2\Rv$ \cite{Fily2014,Fily2016}. This limit can be understood by noting that when a self-propelled filament collides with the vesicle surface it is reoriented into the surface tension plane. When $\lpDyn \gtrsim 2\Rv$, rotation of the filament away from the surface occurs more slowly than realignment with the surface due to subsequent collisions.

The autocorrelation timescale for rotation of a rigid rod scales as $\tCorr \propto a^3$~\cite{Doi2007}. Consistent with this result, measurements on simulations of isolated active filaments give an autocorrelation time of $\tCorr \approx a^3 / 4\pi\gamma$ (dashed line in Fig.~\ref{fig:persistence-time}). The persistence length of motion $\lpDyn$ is therefore
\begin{equation} \label{eq:persistence-length}
    \lpDyn = \tCorr v_0 \approx \frac{\fa a^3}{4\pi m \gamma^2},
\end{equation}
where we have used $v_0 = \fa / m\gamma$ as the active propulsion velocity. Recalling that the system enters the strong confinement limit when this persistent length is on the order of the vesicle diameter, $2\Rv$, the threshold aspect ratio for strong confinement is
\begin{equation} \label{eq:strong-confinement}
    a \gtrsim \left( \frac{8\pi m \gamma^2 \Rv}{\fa} \right)^{1/3}.
\end{equation}
For the activity of Fig.~1, $\fa = 8\kT/\sigma$, the threshold aspect ratio is $\aSC \approx 4.3$.

\subsection{Competition between rotation and collision timescales}

\paragraph{Rotation timescale:}
Consider a rigid, self-propelled rod, consisting of $N + 1$ monomers of mass $m$ and bond length $b$, that is moving toward a flat boundary. The rod is propelled by an active force with magnitude $\fa$ along the rod's tangent $\vu{p}$. At the point of collision with the boundary, the rod will begin to rotate parallel to the wall. The rotation will happen over a timescale equal to the that over which the rod would have moved its own length, $\tR \sim m \gamma L / \fa$, where $L = Nb$ is the rod's length. More explicitly, the (overdamped) equation of motion for the $n$th monomer of the rod is
\begin{equation} \label{eq:overdamped-eom}
    m \gamma \pdv{\vb{r}_n(t)}{t}
        = \fa \vu{p}(t)
        + \vb{N}(s, t),
\end{equation}
where $\vb{r}_n(t) = \vb{R}(t) + nb\vu{p}(t)$ is the position of the $n$th monomer. The last term is the normal force due to the boundary. Note that we have assumed the activity-dominated regime, so thermal forces are negligible.

Without loss of generality, we take the boundary to be oriented such that its normal vector is $\vu{y}$, and the rod to lie in the $x$-$y$ plane. The normal force can then be written as
\begin{equation}
    \vb{N}(s, t) 
        = \fa (\vu{p} \cdot \vu{y}) \left[
            \left(\frac{\vu{p}\cdot\vu{y}}{\vu{p}\cdot\vu{x}}\right) \vu{x}
            - \vu{y}
        \right] \delta_{n,N}.
\end{equation}
Subtracting Eq.~\eqref{eq:overdamped-eom} for $n = 0$ from that of $n = N$ yields
\begin{equation} \label{eq:phat-dynamics}
    m \gamma L \pdv{\vu{p}}{t}
    = \fa (\vu{p} \cdot \vu{y}) \left[
        \left(\frac{\vu{p}\cdot\vu{y}}{\vu{p}\cdot\vu{x}}\right) \vu{x}
            - \vu{y}
        \right].
\end{equation}
The only timescale in this equation is
\begin{equation} \label{eq:rotation-timescale}
    \tR = \frac{m\gamma L}{\fa}.
\end{equation}

\paragraph{Collision timescale:}
To determine the collision timescale, we first must compute the collisional cross section, $\sigmaColl$, of two self-propelled rods. Consider two rigid rods of length $L$ and diameter $\sigma$. One rod is located at $\vb{r}$ with orientation $\vu{p}$, and the other is located at the origin with orientation $\vu{q}$. Again neglecting thermal motion, we assume the rods move in the direction of their tangents with an active velocity $v_0 = \fa / m\gamma$. We choose our reference frame such that the rod at the origin is stationary, and the other moves at the relative velocity $\vb{v} = v_0 (\vb{p} - \vb{q})$. This defines an axis $\vu{w} = \vb{v} / |\vb{v}|$. Projecting the rods onto the plane perpendicular to $\vu{w}$, we find that both rods have the projected length
\begin{equation} \label{eq:projected-length}
    L' = \left(\frac{1 + \vu{p} \cdot \vu{q}}{2}\right) L.
\end{equation}
We can additionally project the rods' orientations on to this plane:
\begin{align}
    \vu{p}' &
        = \frac{\vu{p} - (\vu{p} \cdot \vu{w})\vu{w}}{|\vu{p} - (\vu{p} \cdot \vu{w})\vu{w}|}
        = \frac{\vu{p} + \vu{q}}{|\vu{p} + \vu{q}|} \\
    \vu{q}' &
        = \frac{\vu{q} - (\vu{q} \cdot \vu{w})\vu{w}}{|\vu{q} - (\vu{q} \cdot \vu{w})\vu{w}|}
        = \vu{p}'.
\end{align}
Notably the rods are always parallel when projected into this plane. Treating the projected shapes of the rods as simple rectangles of length $L'$ and width $\sigma$, the collision region is also a rectangle, with side lengths $2L'$ and $2\sigma$, giving a collision area of
\begin{equation}
    A(\vu{p}, \vu{q}) = 4L'\sigma = 2L\sigma (1 + \vu{p} \cdot \vu{q}).
\end{equation}
This approximation of the projected rod shape is valid in the long rod limit, $L \gg \sigma$, since accounting for the exact shape of the ends of the rods will only contribute a term of order $\mathcal{O}(\sigma^2)$. Averaging over all orientations $\vu{p}$ and $\vu{q}$ leads to the total collisional cross section:
\begin{equation} \label{eq:collision-cross-section}
    \sigmaColl = \int A(\vu{p}, \vu{q}) \frac{\dd{\vu{p}}}{4\pi} \frac{\dd{\vu{q}}}{4\pi} = \frac{2}{\pi} L \sigma.
\end{equation}
If the number density of rods is $\rho$, then the collision timescale is defined by the relation
\begin{equation}
    \rho \sigmaColl (L + v_0 \tC) = 1.
\end{equation}
That is, a rod will explore a volume $V(t) = \sigmaColl (L + v_0 t)$ in time $t$, and therefore will collide with $\rho V(t)$ rods on average in that time. The collision timescale is the time $\tC$ such that it will have collided with one rod on average. Thus,
\begin{equation} \label{eq:collision-timescale}
    \tC = \frac{1}{\rho\sigmaColl v_0} - \frac{L}{v_0}
\end{equation}

\paragraph{Competition of timescales:}
Recall the rotation timescale $\tR = L / v_0$. The collision timescale can be written in terms of $\tR$ as
\begin{equation}
    \tC = \left(\frac{1}{\rho\sigmaColl L} - 1\right)\tR,
\end{equation}
We expect that highly-aligned configurations, such as polar rings and caps, will form when $\tC \lesssim \tR$. The number density is related to the volume fraction as $\rho = 4 \phi / \pi L \sigma^2$ (treating the rods as cylinders). Thus, we expect a transition provided $2 \rho \sigma L \gtrsim 1$, or
\begin{equation} \label{eq:capping-transition}
    \phi a \gtrsim (\pi / 4)^2,
\end{equation}
where $a = L / \sigma$ is the rod aspect ratio. Finally, we note that performing the same calculation in 2D results in the same scaling.

\subsection{Calculation of number of caps}

\paragraph{Derivation of the effective `free energy':}
We assume that activity induces an effective attractive interaction between active rods, leading rods to preferentially form smectic layers. These smectic layers are sheared by the curvature of the confining vesicle. We assume that there is a force that is linear in displacement that resists this shear, as well as an interfacial tension at the edges of the cap. Using the parameterization shown in Fig.~2b, we align the cap to the $z$-axis and write the height field of the cap as a function of the polar angle $\theta$:
\begin{equation}
    h(\theta) = \Rv \cos \theta.
\end{equation}
Note that we have assumed the cap radius is the same as the undeformed vesicle radius. Active forces may deform the vesicle at the caps, leading to caps having a radius that is different from the vesicle's unperturbed radius. Thus, this calculation will be most accurate when the vesicle is not deformed too much from its initial state (high vesicle stiffness or low activity).

The gradient in the height field gives the local shear energy density, which can be integrated over the cap surface to find the total shear energy, with shear modulus $G \sim \fa$:
\begin{equation} \label{eq:shear-energy}
\begin{aligned}
    u_\text{shear}(\theta)
        &= \frac{1}{2} G \int \dd{S} \left|\pdv{h}{(\Rv \sin \theta)}\right|^2 \\
        &= \pi \Rv^2 G \int_0^\theta \dd{\theta'} \sin\theta' \tan^2\theta' \\
        &= \pi \Rv^2 G \frac{(1 - \cos\theta)^2}{\cos \theta}.
\end{aligned}
\end{equation}
The cap has radius $\Rv \sin \theta$, and therefore the interfacial energy can be written as
\begin{equation} \label{eq:interfacial-energy}
    u_\text{int}(\theta) = 2\pi \Rv \gamma \sin \theta,
\end{equation}
where $\gamma \sim \fa L$ is the interfacial tension.

If the total number of rods in the system is $\Nf$, and we assume that all caps are the same size, then the total number of caps is given by
\begin{equation} \label{eq:total-caps}
    \nCap = \frac{\Nf}{2\pi \Rv^2 \rho (1 - \cos \theta)},
\end{equation}
where $\rho = 2 / \sqrt{3} \sigma^2$ is the hexagonal close packing density, and $\sigma$ is the rod diameter. Therefore, the total energy (relative to a reference state corresponding to a perfectly aligned smectic layer of rods) is
\begin{equation}
\begin{aligned}
    U(\theta)
        &= \nCap(\theta) [u_\text{shear}(\theta) + u_\text{int}(\theta)] \\
        &= \frac{G N}{\rho (1 - \cos \theta)} \left[
            \frac{(1 - \cos\theta)^2}{2\cos\theta}
            + \frac{\gamma}{G \Rv} \sin\theta
        \right].
\end{aligned}
\end{equation}
From this, we define the (normalized) free energy per filament $f(\theta)$ as
\begin{equation} \label{eq:free-energy}
    f(\theta)
    = \frac{\rho U(\theta)}{GN}
    = \frac{1}{1 - \cos \theta} \left[
        \frac{(1 - \cos\theta)^2}{2\cos\theta}
        + \zeta \sin\theta
    \right],
\end{equation}
where $\zeta = \gamma / G \Rv = c L / \Rv$, with the proportionality constant $c$ the only free parameter.
\\

\paragraph{Computing the number of caps:}
Using Eq.~\eqref{eq:free-energy}, we first compute the optimal $\theta$ as
\begin{equation} \label{eq:optimal-theta}
    \theta^* = \argmin_\theta f(\theta)
\end{equation}
and then evaluate $\nCap(\theta^*)$. This yields the number of caps as a function of the parameter $\zeta$, which we can fit to our simulation data.

Provided $\zeta$ is small, we can find an asymptotic solution in the following way. Let $z = \cos \theta / (1 - \cos \theta)$ for convenience. Since $z(\theta)$ is a monotonic function of $\theta$ (for $0 \leq \theta < \pi$), minimizing Eq.~\eqref{eq:free-energy} with respect to $z$ is equivalent to minimizing with respect to $\theta$. The free energy can then be written as
\begin{equation}
    f(z) = \frac{1}{2z} + \zeta \sqrt{1 + 2z}.
\end{equation}
Taking the derivative and setting it to zero, we obtain the following equation for $z$:
\begin{equation}
    \frac{\zeta}{\sqrt{1 + 2z}} - \frac{1}{2z^2} = 0.
\end{equation}
Squaring both sides, we can rearrange this into the form
\begin{equation}
    \frac{z^4}{1 + 2z} = \frac{1}{4\zeta^2}.
\end{equation}
If $\zeta \ll 1$, then the right hand side is very large and so we can assume that $z \gg 1$ as well. Approximating $1 + 2z \approx 2z$, this reduces to
\begin{equation} \label{eq:z-solution}
    z \approx (2\zeta^2)^{-1/3}
\end{equation}
or, equivalently,
\begin{equation} \label{eq:theta-solution}
    \cos \theta \approx \frac{1}{1 + (2\zeta^2)^{1/3}}.
\end{equation}
Recall that we expect $\zeta \propto L / \Rv$. We can instead write $\zeta$ as a ratio of the rod aspect ratio $a$ to a critical aspect ratio, $\zeta = a/a^*$, with $a^* \cong \Rv/\sigma \gg 1$. Inserting Eq.~\eqref{eq:theta-solution} into Eq.~\eqref{eq:total-caps}, and absorbing constants into $a^*$, we obtain the result shown in the main text
\begin{equation}
    \nCap = A \phi a^{-1} [1 + (a / a^*)^{-2/3}],
\end{equation}
where we have used $N = \phi \Vv / \Vf$, with $\Vv = 4 \pi \Rv^3/3$ and $\Vf = \pi L \sigma^2 / 4$, and $A = 8\Rv / 3 \pi \rho \sigma^3$.

\textit{Dependence of $a^*$ on activity:} We anticipate that the critical aspect ratio may depend on activity for several reasons. First, the critical aspect ratio will depend on the \textit{local} vesicle curvature in the vicinity of the cap, $a^*\cong\Rv^\text{local}/\sigma$, with $\Rv^\text{local}\le\Rv$ because the active force from the rods will locally deform the vesicle from its unperturbed curvature. The extent of deformation will increase with activity. Second, at very low activity $\fa\lesssim\kT/\sigma$ (or $\fa\lesssim1$ in our units), thermal motions become relevant and the caps are not as well-formed (see Fig.~\ref{fig:ragged_caps} and SI Movie 4).

\subsection{Onset of dynamic caps}


Assuming all $\Nf$ rods are contained within a cap, the total area occupied by rods is
\begin{equation}
    \Atail = \Nf / \rho,
\end{equation}
where, as before, $\rho = 2/\sqrt{3} \sigma^2$ is the hexagonal close packing density. Now, let us consider a state with all rods organized into caps, and assume that the vesicle is approximately spherical with radius $\Rv$. Then, the tails of the rods will be located approximately on the surface of an inner spherical region, with radius $R = \Rv - \Lrod$. If the total area of rod tails $\Atail$ occludes a large fraction of the surface area of this inner region, then the tails of the rods are likely to interact. Such interactions can disrupt the stability of the caps, potentially leading to motions and/or breakup of caps. This occurs when
\begin{equation}
    \Nf / \rho \gtrsim 4\pi (\Rv - \Lrod)^2
    \to \Nf \gtrsim 4\pi \rho \Rv^2 (1 - \Lrod / \Rv)^2.
\label{eq:motileCaps}
\end{equation}
Based on this analysis, caps will become motile above threshold values of the length and number of filaments given by Eq.~\eqref{eq:motileCaps}. 

We can recast these results in terms of the volume fraction $\phi$ and aspect ratio $a$ of the filament. Using $\Nf \sim \phi / a$, we write
\begin{equation}
    \phi \gtrsim C a (1 - a\sigma / \Rv)^2,
\end{equation}
where $C \propto \Lrod / \Rv$. Consistent with this result, our simulation results show that motile caps arise in the lower-right region of the parameter space of Fig.~1.

\subsection{Effects of filament semiflexibility}
\label{sec:semiflexible}

As we show in Sec.~A, the filament orientation correlation timescale is
\begin{equation}
    \tCorr = \frac{a^3}{4\pi \gamma}.
\end{equation}
We assume that for semiflexible rods we can make the substitution $a \to \lp / \sigma$, where $\lp$ is the filament persistence length. Further, simulations show~\cite{Joshi2019}
\begin{equation}
    \lp = \frac{2 \kappa b / \kT}{1 + (\fa \sigma / \kT)^2},
\end{equation}
where $b = \sigma / 2$ is the bond length between filament monomers and $\kappa$ is strength of the harmonic angle potential that controls the semiflexibility of the filament. Assuming that $\fa$ is large, we find
\begin{equation} \label{eq:renormalized-lp}
    \lp/\sigma \approx \frac{\kT}{\sigma^2} \frac{\kappa}{\fa^2}.
\end{equation}
The system enters the strong confinement limit when $\tCorr \gtrsim 2\Rv / v_0$, where $v_0 = \fa / m\gamma$ is the filament propulsion velocity. Using Eq.~\eqref{eq:renormalized-lp}, this can be rearranged to find
\begin{equation} \label{eq:faSC}
    \faSC \sigma/ 
    \kT \gtrsim C (\kappa / \kT)^{3/5},
\end{equation}
where $C = \left(\frac{\kT}{8 \pi m \gamma^2 \sigma \Rv}\right)^{1/5}$. This is shown as a dashed line in Fig.~5.

Note that there are no fitting parameters used in this analysis.

\subsection{Equations of motion and simulation details}

We simulate a vesicle with $\Nv = 2432$ monomers and a nominal radius of $\Rv \approx 25\sigma$, measured as the distance from the center of mass of the vesicle to the center of any given monomer within the vesicle in its undeformed state. For this radius, we set the vesicle monomer diameter to $\sigmaVes = a\sigma$ with $a \approx 1.934$ to ensure that there were no holes in the vesicle that the active filaments could escape from. A filament is a linear chain of $M$ monomers, each with diameter $\sigma$. The vesicle is filled with $\Nf$ such filaments, with $\Nf$ varied to control the volume fraction $\phi=\Nf \Vf / \Vv$. See Fig.~\ref{fig:model}.

Bonded monomers in both the vesicle and the filaments interact through an expanded FENE potential:
\begin{equation} \label{eq:fene-potential}
    U_\text{FENE}(r)
    = -\frac{1}{2} \kStretch \Delta^2 \ln \left[
        1 - \left( \frac{r - b}{\Delta} \right)^2
    \right],
\end{equation}
where $r$ is the distance between the two bonded monomers, $\kStretch$ is the bond strength, $b$ is the preferred bond length, and $\Delta$ is the maximum deviation; that is, $|r - b| < \Delta$. For the vesicle, we set $\kStretch = 1000\kT/\sigma^2$, $\bves \approx 1.934\sigma$ (for hexamer bonds) or $\bpent \approx 1.645\sigma$ (for pentamer bonds), and $\Delta = b/2$. For the active filaments, we set $K = 2000\kT/\sigma^2$, $\bfil = 0.5\sigma$, and $\Delta = 0.4\sigma$. These stiff bond potentials act to nearly constrain the length and area of the filaments and vesicle, respectively.

To penalize curvature, neighboring triangles on the vesicle (those that share an edge) also interact through a harmonic dihedral potential. Each triangle $i$ defines a unique normal vector $\vu{n}_i$. For neighboring triangles $i$ and $j$, the interaction potential is given by
\begin{equation} \label{eq:dihedral-potential}
    U_\text{dih}(\phi) = \kDih (1 - \cos \phi),
\end{equation}
where $\cos \phi = \vu{n}_i \cdot \vu{n}_j$. For most simulations in the main text, we set $\kVesBend = 5000\kT$ so that the vesicle has a relatively large bending modulus $\kappa = \sqrt{3} \kVesBend$~\cite{Gompper1996}. Infinitely rigid vesicles are implemented by not integrating the equations of motion for the vesicle monomers so that they are always in their initial (spherical) configuration.

Filament semiflexibility is incorporated through a harmonic angle potential of the form:
\begin{equation} \label{eq:angle-potential}
    U_\text{angle}(\theta) = \kFilBend \theta^2,
\end{equation}
where $\theta$ is the angle between two neighboring bonds $i$ and $j$; that is, $\cos \theta = \vu{b}_i \cdot \vu{b}_j$, where $\vu{b}_{\{i,j\}}$ are neighboring bond vectors.

Non-bonded monomers interact sterically through an expanded WCA potential given by
\begin{equation} \label{eq:wca-potential}
    U_\text{WCA}(r) = 4 \epsilon \left[
        \left(\frac{\sigma}{r - \Delta}\right)^{12}
        - \left(\frac{\sigma}{r - \Delta}\right)^{6}
    \right]
    \quad r < r_c,
\end{equation}
where $r$ is the distance between two interacting monomers, $\epsilon$ is the strength of the interaction, $\Delta$ shifts the potential, and $\rc = 2^{1/6}\sigma + \Delta$ is a cutoff distance. For interactions between two filament monomers, we set $\Delta = 0$, and for interactions between a filament monomer and a vesicle monomer we set $\Delta = \sigmaVes / 2$. This ensures that active filaments cannot escape the vesicle. In all cases, we set $\epsilon = \kT$. We do not consider excluded volume interactions between two vesicle monomers as the stiff bond and dihedral potentials act to keep vesicle monomers far apart.

Given these interactions, the total energy is the sum
\begin{equation} \label{eq:total-energy}
\begin{split}
    U_\text{tot} =
    & \sum_{\text{pairs}} U_\text{WCA}(r)
    + \sum_{\text{bonds}} U_\text{FENE}(r) \\
    & + \sum_{\text{angles}} U_\text{angle}(\theta)
    + \sum_{\text{dih}} U_\text{dih}(\phi).
\end{split}
\end{equation}
The equation of motion for any given atom $i$ within molecule $\alpha$ is then given by
\begin{equation} \label{eq:equation-of-motion}
\begin{split}
    m \pdv[2]{\vb{r}_i^\alpha}{t} =
    &-\pdv{U_\text{total}}{\vb{r}_i^\alpha} - m \gamma \pdv{\vb{r}_i^\alpha}{t} \\
    &+ (1 - \delta_{\alpha,0}) \fa
        \frac{\vb{r}_{i+1}^\alpha - \vb{r}_{i-1}^\alpha}{|\vb{r}_{i+1}^\alpha - \vb{r}_{i-1}^\alpha|}
    + \vb{\xi}_i^\alpha(t).
\end{split}
\end{equation}
Here, $\alpha = 0$ represents the vesicle, while $\alpha > 0$ represents a filament. The first term on the right hand side consists of the conservative forces present in the system, the second term damps the motion through a viscous force (with the damping coefficient set to $\gamma = 1/\tau$), the third term applies an active force to the atoms of each filament along the tangent, and the last term adds random thermal forces. The thermal forces are modeled as Gaussian white noise with moments
\begin{equation} \label{eq:thermal-moments}
\begin{split}
    \ev{\vb{\xi}_i^\alpha(t)} &= 0, \qand \\
    \ev{\vb{\xi}_i^\alpha(t) \cdot \vb{\xi}_j^\beta(t')} &=
        6 \gamma \kT \delta_{ij} \delta_{\alpha \beta} \delta(t - t').
\end{split}
\end{equation}
Simulations are run with a timestep of $\delta t = 10^{-3}\tau$ for a total time of $10^4 \tau$. For the first two state diagrams shown in the main text (Fig.~1 and Fig.~3) we ran 9 trials for each set of parameters, while for the last two (Fig.~5 and Fig.~6) we only ran one trial.

\subsection{State detection algorithm}

We classify states by a combination of the vesicle conformation and the organization of the rods within the vesicle. The classifications are as follows
\begin{itemize}
    \item \textit{undeformed/spherical}: the vesicle is in a nearly spherical state (symbolized by $\bullet$)
    \item \textit{deformed/other}: the vesicle is in a deformed state, but the rod organization is not consistent or difficult to detect; typically, this occurs for states in which the filaments form unstable polar rings that form and break apart over time (symbolized by $\star$)
    \item \textit{oblate/ring}: the vesicle is an oblate spheroid, and the filaments are organized into a stable polar ring (symbolized by $\circ$)
    \item \textit{capped}: the filaments are organized into stable caps, with varying amounts of vesicle deformation (symbolized by $\circ$ with the number of intersecting lines in the symbol equal to the median number of detected caps)
    \item \textit{dynamic capped}: the filaments are organized into caps that are dynamic and/or motile (symbolized by $\diamond$)
\end{itemize}
Our classification algorithm is constructed as follows. We first classify the vesicle conformation. We define the asphericity $\alpha$ of the vesicle to be
\begin{equation} \label{eq:asphericity}
    \alpha = \frac{1}{6\sqrt{\pi}} \frac{S^{3/2}}{V},
\end{equation}
where $S$ is the surface area of the vesicle and $V$ is the volume. Note that $\alpha = 1$ for a sphere, and $\alpha > 1$  for any non-spherical shape. Additionally, we measure the gyration tensor $G$ as
\begin{equation}
    G = \frac{1}{\Nv} \sum_{i=1}^{\Nv} \vec{r}_{i} \otimes \vec{r}_{i},
\end{equation}
where $\Nv$ is the number of monomers making up the vesicle, and $\vec{r}_i$ is the position of the $i$th monomer. Let $a$, $b$, and $c$ be the eigenvalues of $G$ such that $a \leq b \leq c$. Then define
\begin{align*}
    x &= (b - a) / (b + a) \\
    y &= (c - b) / (c + b).
\end{align*}
Using $\alpha$, $x$, and $y$, we classify the vesicle conformation into four categories:
\begin{itemize}
    \item \textit{spherical}: $\alpha \leq 1.01$ and $x \leq 0.1$ and $y \leq 0.1$,
    \item \textit{oblate}: $x > 0.1$ and $y \leq 0.1$,
    \item \textit{prolate}: $x \leq 0.1$ and $y > 0.1$,
    \item \textit{non-spherical}: otherwise.
\end{itemize}
Next, we attempt to classify the filament organization.

First, we determine if there are any caps in the system. To do this, we compute two fields on the vesicle surface: a polarization field $\vb{p}_i$ and an order parameter $q_i$ that measures how perpendicular the filaments are to the vesicle, both of which are evaluated at each vesicle monomer $i$. Let $\vu{n}_i$ be the unit normal vector of the vesicle at the $i$th vesicle monomer, and let $M_i$ be the set of filament monomers that are within $4\sigma$ of the $i$th vesicle monomer. Then
\begin{equation}
    \vb{p}_i = \frac{\sum_{j \in M_i} \vu{t}_j}{\left|\sum_{j \in M_i} \vu{t}_j\right|},
\end{equation}
where $\vu{t}_j$ is the unit tangent vector of the filament evaluated at filament monomer $j$. Similarly,
\begin{equation}
    q_i = \frac{1}{|M_i|} \sum_{j \in M_i} \vu{t}_j \cdot \vu{n}_i
\end{equation}
To find potential caps, we first find the vesicle monomer $i^*$ such that $q_{i^*}$ is maximized. We then use the procedure described in Algorithm~\ref{alg:get-cap} to find a cap based on this seed monomer. To find additional caps, we choose a new seed point by finding another monomer $j^*$ such that $q_{j^*}$ is the maximal value among all monomers that are not yet a member of a cap. We continue this process until all vesicle monomers have been visited, or until we do not find any additional caps. Once all candidate caps are found, we discard any caps that comprise fewer than $\lceil \sqrt{\Nv} \rceil = 50$ monomers.

\normalem
\IncMargin{1em}
\begin{algorithm*} \label{alg:get-cap}
\caption{Algorithm to determine vesicle monomers belonging to a cap}

\SetKwInOut{Input}{input}
\SetKwInOut{Output}{output}

\SetKwData{cap}{cap}
\SetKwData{todo}{todo}
\SetKwData{visited}{visited}

\SetKwFunction{pop}{pop}
\SetKwFunction{push}{push}
\SetKwFunction{append}{append}
\SetKwFunction{insert}{insert}

\Input{A seed vesicle monomer $i^*$}
\Output{A list of vesicle monomers belonging to a cap}
\BlankLine
initialize \cap to empty list\;
initialize \visited to empty set\;
initialize \todo to stack containing only $i^*$\;
\While{\todo is not empty}{
    $i \gets$ value popped from \todo\;
    insert $i$ into \visited\;
    \If{$\vu{p}_i \cdot \vu{p}_{i^*} \geq \cos(\pi/6)$} {
        append $i$ to \cap\;
        $N_i \gets$ set of monomers neighboring monomer $i$\;
        \For{$j \in N_i$} {
            \If{$j$ not in \visited and $\vu{p}_j \cdot \vu{p}_{i^*} \geq \cos(\pi/6)$}{
                push $j$ on to \todo\;
            }
        }
    }
}

\end{algorithm*}
\DecMargin{1em}
\ULforem

Finally, as done for the vesicle itself, we compute the smallest and largest eigenvalues, $\lambda_\text{min}$ and $\lambda_\text{max}$ respectively, of the gyration tensor computed using only the filament monomers. Let $z = 1 - \lambda_\text{min} / \lambda_\text{max}$. Then we classify the filament organization as one of the following:
\begin{itemize}
    \item \textit{capped}: if there are a non-zero number of caps,
    \item \textit{polar ring}: if there are no caps, and $z > 0.75$,
    \item \textit{other}: otherwise.
\end{itemize}
The class \textit{other} is very broad, and can include a range of filament organizations and behaviors such as polar flocks on the vesicle surface, or transient polar rings that continuously form and break apart. Usually these states do not result in any large scale deformation of the vesicle, but in some cases the transient ring states can lead to measurable deformation.

For each simulation, we take 11 samples from the last 10\% of the trajectory and measure the vesicle conformation and filament organization at each sample. If we also have multiple trials, then we aggregate these identifications. Finally, we assign an conformation and organization based on the most common measurement.

Having information on both the vesicle conformation and filament organization, we can revisit our earlier classification scheme:
\begin{itemize}
    \item \textit{undeformed/spherical}: the filament organization is \textit{other}, and the vesicle conformation is \textit{spherical},
    \item \textit{deformed/other}: the filament organization is \textit{other}, and the vesicle conformation is not \textit{spherical},
    \item \textit{oblate/ring}: the filament organization is \textit{polar ring}, and the vesicle conformation is \textit{oblate},
    \item \textit{capped}: the filament organization is \textit{capped},
    \item \textit{dynamic capped}: the filament organization is \textit{capped}, but the number of caps is inconsistent.
\end{itemize}
For the dynamic capping states, we classify the number of caps as inconsistent if the majority of states are capped, but no particular number of caps forms a majority of these states.

\subsection{Movie Descriptions}

\begin{itemize}
    \item \texttt{movie-1.mp4}: Animations corresponding to the data shown in Fig.~1. Each element of the figure corresponds to an independent simulation performed at the indicated parameter values, with the volume fraction $\phi$ increasing along rows, and the aspect ratio $a$ increasing along the columns. Other parameters are $\fa = 8$ and $\kFilBend = 10^4$.
    \item \texttt{movie-2.mp4}: Animations corresponding to the data shown in Fig.~3, with the volume fraction $\phi$ increasing along rows, and the active force strength $\fa$ increasing along the columns. Other parameters are $a = 10.5$ and $\kFilBend = 10^4$.
    \item \texttt{movie-3.mp4}: Animations corresponding to the data shown in Fig.~S2, with the filament stiffness $\kFilBend$ increasing along rows, and the active force strength $\fa$ increasing along the columns. Other parameters are $\phi = 0.20$ and $a = 10.5$.
    \item \texttt{movie-4.mp4}: Animations of simulations performed at $\fa \in \{1, 2, 3\}$ for $\phi = 0.2$, $a = 10.5$, and $\kFilBend = 10^4$. These videos show that decreasing the activity leads to smaller vesicle deformations, more ragged caps, and more frequent rod dissociation from caps.
    \item \texttt{movie-5.mp4}: Animations of simulations performed at $\phi \in [0.01, 0.10]$ and $\kVesBend \in [10^2, 10^4]$ for $\fa = 8$, $\phi = 0.1$, $a = 10.5$, and $\kFilBend = 10^4$. These videos show the faceting transition of the vesicle as the rigidity is reduced. Caps are still readily formed and stable, but the polar rings tend to trace a path between facets which destabilizes them.
    \item \texttt{movie-6.mp4}: Animation of a simulation performed at $\fa = 8$, $\kFilBend = 10^4$, $\phi = 0.05$ and $a = 25.5$ in an infinitely rigid vesicle. This video shows that the polar rings that form within a rigid vesicle are unstable, and will break apart and reform repeatedly. In contrast, these parameters lead to a stable polar ring when the vesicle has finite rigidity.
\end{itemize}

\bibliographystyle{apsrev4-1}
\bibliography{references}